\documentclass[letterpaper,11pt]{article}
\pdfoutput=1 % if your are submitting a pdflatex (i.e. if you have
\usepackage{jheppub} % for details on the use of the package, please
\usepackage[utf8]{inputenc}
\usepackage[T1]{fontenc} % if needed
\usepackage[dvipsnames]{xcolor}
\usepackage{multirow}
\usepackage{slashed}
\usepackage[normalem]{ulem}
\usepackage{sidecap}
\usepackage{mathtools}
\usepackage{caption}
\usepackage{subcaption}
\usepackage[toc]{appendix}

\title{Testing freeze-in with axial and vector $Z^\prime$ bosons}

\author[a,b]{Catarina Cosme}
\author[a]{, Ma\'ira Dutra}
\author[a]{, Stephen Godfrey}
\author[a]{, Taylor Gray}

\affiliation[a]{Ottawa-Carleton Institute for Physics,
Carleton University, 1125 Colonel By Drive, Ottawa, Ontario K1S 5B6, Canada}
\affiliation[b]{Instituto de F\'isica Corpuscular (IFIC), Universidad de Valencia, C/ Catedr\'atico Jos\'e Beltr\'an 2, E-46980 Paterna, Spain}

\emailAdd{catarina.cosme@ific.uv.es}
\emailAdd{mdutra@physics.carleton.ca}
\emailAdd{godfrey@physics.carleton.ca}
\emailAdd{taylorgray@cmail.carleton.ca}

\usepackage{calligra}
\usepackage{calrsfs}
\DeclareMathAlphabet{\mathcalligra}{T1}{calligra}{m}{n}
\DeclareMathAlphabet{\pazocal}{OMS}{zplm}{m}{n}

\def\Lag{\pazocal{L}}
\def\M{\pazocal{M}}
\def\to{\rightarrow}
\def\tob{\leftrightarrow}
\def\zp{{Z^\prime}}
\def\figureautorefname~#1\null{Fig.\,#1\null}
\def\tableautorefname~#1\null{Tab.\,#1\null}

\def\equationautorefname~#1\null{Eq.\,(#1)\null}

\abstract{
The freeze-in production of Feebly Interacting Massive Particle (FIMP) dark matter in the early universe is an appealing alternative to the well-known -- and constrained -- Weakly Interacting Massive Particle (WIMP) paradigm. Although challenging, the phenomenology of FIMP dark matter has been receiving growing attention and is possible in a few scenarios. In this work, we contribute to this endeavor by considering a $Z^\prime$ portal to fermionic dark matter, with the $Z^\prime$ having both vector and axial couplings and a mass ranging from MeV up to PeV. We evaluate the bounds on both freeze-in and freeze-out from direct detection, atomic parity violation, leptonic anomalous magnetic moments, neutrino-electron scattering, collider, and beam dump experiments. We show that FIMPs can already be tested by most of these experiments in a complementary way, whereas WIMPs are especially viable in the $Z^\prime$ low mass regime, in addition to the $\zp$ resonance region. We also discuss the role of the axial couplings of $Z^\prime$ in our results. We therefore hope to motivate specific realizations of this model in the context of FIMPs, as well as searches for these elusive dark matter candidates.}

\begin{document}

\today
\maketitle
\flushbottom

\section{Introduction}
\label{sec:intro}

One of the most intriguing puzzles of Particle Physics and Cosmology is the nature of dark matter (DM), a non-relativistic matter component that makes up about $27\%$ of the current cosmic energy \cite{Aghanim:2018eyx}. 
The observational evidence for the existence of DM is overwhelming, relying on its gravitational interaction with ordinary matter. DM is required to explain the anisotropies of the Cosmic Microwave Background (CMB), the flatness of galaxy rotation curves, and the large-scale structure of the Universe. 
Nevertheless, despite the large number of DM candidates
that emerge from theories beyond the Standard Model (SM), the nature of DM remains unknown (see Ref. \cite{Bertone:2004pz}
for a review). Moreover, although many experimental searches have been carried out over the past decades, DM has evaded detection. 

Among the DM candidates, weakly interacting massive particles (WIMPs) are certainly the most popular ones. These particles
were kept in thermal equilibrium with the SM bath in the early Universe, and just when their interaction rates fell below the Hubble rate -- meaning that the interactions could not
keep up with the expansion of the Universe -- they decoupled from the
cosmic bath and its abundance \textit{froze-out}, yielding the observed DM
relic density \cite{Gondolo}. Since the cross-sections that generate the observed
WIMP abundance are typically at the electroweak scale, currently being probed at direct, indirect detection and collider experiments, WIMP models deserve to be completely explored. Although WIMPs dominate the searches for dark matter and might soon be discovered, the current situation of stringent constraints \cite{Baer:2014eja,Arcadi:2017kky} motivate us to look for alternative scenarios.

The tight experimental constraints on dark matter can be evaded by relaxing the assumption of thermal production, provided that the couplings are very small. In this case, DM is produced by the \textit{freeze-in} mechanism, where the abundance is generated by decays and annihilations of SM bath particles into DM \cite{Giudice:2000ex,McDonald:2001vt,Hall:2009bx,Bernal:2017kxu}. These DM candidates are known as feebly interacting massive particles (FIMPs). Although FIMP scenarios are, generically, more difficult to test due to the smallness of the couplings, some can be probed through colliders \cite{Co:2015pka,Calibbi:2018fqf,Belanger:2018sti,No:2019gvl,Heeba:2019jho,Okada:2020cue}, direct \cite{Chu:2011be,Essig:2011nj,Essig:2015cda,Hambye:2018dpi,Bernal:2018ins,Heeba:2019jho,Chang:2019xva,An:2020tcg} and indirect \cite{Brdar:2017wgy,Biswas:2019iqm,Cosme:2020mck} detection searches, as well as astrophysical/cosmological observations \cite{Hambye:2018dpi,Bernal:2018ins,Chang:2019xva,No:2019gvl,Huo:2019bjf}.

An interesting possibility to connect the dark and visible sectors is through a new gauge boson $Z^\prime$, associated with  a new gauge $U(1)$ symmetry. They can arise in the
%well-motivated 
context of supersymmetric \cite{Cvetic:1998jxa,Chun:2008by,Frank:2020byg}, GUT \cite{London:1986dk,Hewett:1988xc,Arcadi:2017atc}, and string-inspired models \cite{Cvetic:1995rj,Cleaver:1998gc}, or be phenomenologically invoked from a bottom-up perspective (see Ref. \cite{Langacker:2008yv} for a review on heavy $Z^\prime$'s). 
Moreover, they profit from dedicated searches, complemented by the searches for dark matter \cite{Zhu:2007zt,Frandsen:2011cg,An:2012ue,Arcadi:2013qia,Hooper:2014fda,Altmannshofer:2014pba,Alves:2016cqf,Arcadi:2017hfi,Albert:2018jwh,Blanco:2019hah,Belanger:2020gnr,Okada:2020cue,Cadeddu:2020nbr,Frank:2020byg,Diener:2011jt,Aad:2019fac,Sirunyan:2019vgt,Schael:2013ita,Lees:2014xha,
Lees:2017lec,Bennett:2004pv,
Bennett:2006fi}. 

The special case of a pure vector $\zp$ which kinetically mixes to photons, the dark photon, has been extensively studied in the literature, both in the context of WIMPs \cite{Pospelov:2007mp,Dutra:2018gmv,Cho:2020mnc,Filippi:2020kii,Fabbrichesi:2020wbt,Bernreuther:2020koj} and FIMPs \cite{Heeba:2019jho,Chu:2011be,Hambye:2018dpi,Chang:2019xva}. It is already known that sub-GeV dark photons have the interesting feature of enhancing direct detection rates and rendering FIMPs testable  \cite{Hambye:2018dpi}.

In this work, we intend to further investigate the phenomenology of FIMPs in a $\zp$ portal model as well as access the viable WIMP parameter space in a wider $\zp$ mass range, from MeV to PeV. We therefore study the freeze-out and the freeze-in production of a Dirac fermion DM candidate, $\chi$, which only interacts with the SM through a $Z^\prime$ boson. We consider both vector and axial-vector couplings in the $\zp$ currents, for a wide range of values. We evaluate how the region of our parameter space providing the right amount of DM is constrained by direct detection (XENON1T), colliders (LHCb, ATLAS, LEP II, BaBar), neutrino-electron scattering, atomic parity violation, electron and muon anomalous magnetic moments, and electron beam dump experiments (E137, E141).

In the presence of axial couplings, direct detection bounds are easily evaded and would hardly probe the FIMP scenario. Nevertheless, we show that most of the experiments considered in this work are already testing the parameter space of FIMPs. As a consequence, even though axial $\zp$'s are tightly constrained, especially in their light regime \cite{Fayet:2007ua,Bouchiat:2004sp}, they provide a viable framework for FIMP dark matter.

The paper is organized as follows: in \autoref{sec:models}, we introduce our model. In \autoref{sec:relic density}, we describe how the present DM abundance can be achieved both in the context of the freeze-out and the freeze-in mechanisms, whereas in \autoref{sect:constraints} we show how the parameter space can be constrained using information from various experiments, as well as cosmological and theoretical bounds. Finally, in \autoref{sec:Results} we present our main results and in \autoref{sec:conclusion}, we conclude.

\section{The model}
\label{sec:models}

In this work, we address the possibility of testing feebly interacting dark matter whose interactions with ordinary matter are mediated by a $\zp$, the massive gauge boson coming from an extra gauge $U(1)$ symmetry, $U(1)^\prime$. 

We consider an extra Dirac fermion, $\chi$, as our dark matter candidate. When both the SM fermions ($f$) and $\chi$ are charged under $U(1)^\prime$, the relevant Lagrangian is given by
\begin{equation}\label{Eq:lagrangian}
\Lag \supset  - m_\chi \bar \chi \chi - \frac{1}{2}m_\zp Z^\prime_\mu {Z^\prime}^\mu + \bar \chi \gamma^\mu (V_\chi-A_\chi \gamma_5)\chi Z^\prime_\mu + \sum_f  \bar f \gamma^\mu (V_f-A_f \gamma_5) f Z^\prime_\mu\, ,
\end{equation}
where $m_\chi$ and $m_\zp$ are the dark matter and $\zp$ masses, and $V_{\chi,f}$ and $A_{\chi,f}$ are respectively vector and axial dimensionless couplings. 
These are the six free parameters considered in our analysis. 
In our convention, in terms of the dark gauge coupling and chiral charges, we have $V_f=g_\zp({X_f}_L+{X_f}_R)/2$ and $A_f=g_\zp({X_f}_L-{X_f}_R)/2$. Because SM neutrinos are left-handed, we will always fix $V_\nu = A_\nu$. 

For simplicity, we neglect possible mass or kinetic mixing between the SM hypercharge gauge boson and the $\zp$. Note that kinetic mixing can also be generated at loop level via SM fermions. However, given that the kinetic mixing has been constrained to be small \cite{Williams:2011qb}, these small corrections do not qualitatively change our results.

As we consider in \autoref{sect:constraints}, many experiments search for dark matter and $\zp$'s. In order to evade the current stringent bounds while still hoping to discover them in upcoming experiments, many realizations of a $\zp$ portal were proposed.

Direct detection experiments are less sensitive to dark matter in certain $\zp$ models. For \textit{purely axial} $\zp$'s, with $A_f \neq 0$ and $V_f = 0$ (which means ${X_f}_L = - {X_f}_R$ for all fermions) \cite{Lebedev:2014bba,Hooper:2014fda,Alves:2016cqf,Ismail:2016tod,Casas:2019edt}, the scattering between dark matter and quarks is mainly spin-dependent, which is less constrained\footnote{Spin-independent scatterings might still be loop-induced and dominate certain regions of the parameter space if $V_\chi \neq 0$ \cite{Alves:2016cqf}.}. Furthermore, if we had chosen $\chi$ as a Majorana fermion (neutral), its vector current, associated with its charge under $U(1)^\prime$, would have to be exactly zero. It is also interesting to notice that potential signals of new physics can be explained by considering non-vanishing axial couplings, as in the case of MeV anomalies \cite{Kahn:2016vjr} and the Galactic Center gamma-ray excess \cite{Hooper:2014fda}.

The constraints on $\zp$ interactions with standard fermions are also quite stringent. Flavour and generation specific realizations are then usually invoked. This is the case of \textit{leptophilic} models, with the $\zp$ coupling mainly to leptons \cite{Foldenauer:2018zrz,Fox:2008kb,Kopp:2009et,Bell:2014tta,Chen:2015tia,DEramo:2017zqw,Duan:2017qwj,Ellis:2018xal,Sadhukhan:2020etu,Buras:2021btx}; \textit{leptophobic} models, where the $\zp$ couples mainly to quarks \cite{Buckley:2011mm,Gondolo:2011eq,An:2012va,Alves:2013tqa,Ellis:2018xal,Frank:2020byg}; and models where the $\zp$ couples to the SM fermions in a \textit{non-universal} way, for instance, with couplings to only the third generation quarks or first generation leptons \cite{Hooper:2014fda,Blanco:2019hah}. Allowing $\zp$'s to decay predominantly into invisible states (in our case, $V_\chi, A_\chi \gg V_f, A_f$) also makes collider bounds weaker. 

Other possibilities are \textit{sequential} $\zp$ models, where $V_f$ and $A_f$ are the same as the SM $Z$ bosons couplings, or simply re-scaled \cite{Langacker:2008yv,Alves:2013tqa,Alves:2015pea,Arcadi:2017hfi}; and \textit{kinetic mixing} portals \cite{Pospelov:2008zw,Mambrini:2010dq,Mambrini:2011dw, Chu:2013jja,Fayet:2016nyc,Gherghetta:2019coi}, coming from the fact that the kinetic terms of two $U(1)$ gauge bosons are in general non-diagonal. The case of a \textit{dark photon} ($A_f = 0$, or $X_{f_L}=X_{f_R}$, for all SM fermions), typically kinetically mixed to photons, has been extensively studied in the literature and is often regarded as a target for future experiments \cite{Dutra:2018gmv,Heeba:2019jho,Cho:2020mnc,Filippi:2020kii,Fabbrichesi:2020wbt,Bernreuther:2020koj}. To motivate model-dependent analyses with specific charge assignments in the context of FIMPs, in what follows we focus on universal $V_f$ and $A_f$ couplings. 

Finally, we would like to stress that realistic and UV complete $\zp$ models with a low-energy Lagrangian as in \autoref{Eq:lagrangian} requires the introduction of additional fields. In the presence of axial couplings, gauge invariance of the Yukawa sector requires the introduction of additional scalars \cite{Kahn:2016vjr}. On the other hand, additional scalars might also be needed to generate the masses of the $\chi$ and $\zp$. Moreover, the presence of axial currents introduce triangle anomalies which must be cancelled. This is typically done by invoking new fermions charged under both standard and dark $U(1)$'s \cite{Hooper:2014fda,Alves:2015mua,Ismail:2016tod}. To keep our analysis as model-independent as possible, we assume that all these additional fields are too heavy to impact the production of dark matter\footnote{Note that, depending on the mechanism for mass generation of the BSM states, such a mass hierarchy might not be easily achieved (see for instance Refs. \cite{Ismail:2016tod,Kahlhoefer:2015bea}).}. Our results would therefore be applicable to any chiral charge assignments under these assumptions.

In the next section, we show how both the strength of the couplings and the masses of $\chi$ and $\zp$ determine the way dark matter is produced in the early universe. 

\section{Relic density}
\label{sec:relic density}

\begin{figure}[!t]
    \centering
    \begin{subfigure}[b]{0.2\linewidth}
     \includegraphics[width=\textwidth]{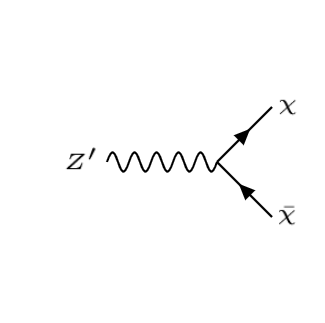}
     \caption{}
    \end{subfigure}
    \begin{subfigure}[b]{0.2\textwidth}
     \includegraphics[width=\textwidth]{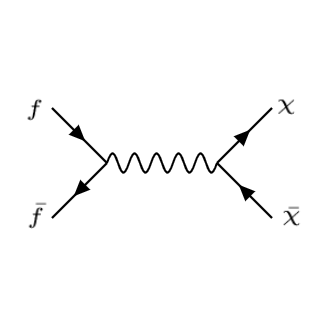}
     \caption{}
    \end{subfigure}
    \begin{subfigure}[b]{0.2\textwidth}
     \includegraphics[width=\textwidth]{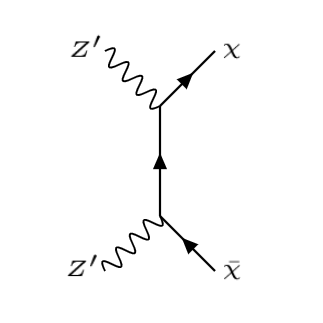}
     \caption{}
    \end{subfigure}
    \begin{subfigure}[b]{0.2\textwidth}
    \includegraphics[width=\textwidth]{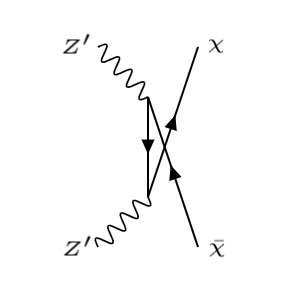}
    \caption{}
    \end{subfigure}
    \caption{The Feynman diagrams for the DM production processes in our model: $Z'$ decays (a), SM fermion annihilations through s-channel exchange of $\zp$ (b), and $\zp$ annihilations through t and u-channel exchanges of $\chi$ (c,d).}
\label{fig:feynman_diagrams}
\end{figure}

Our dark matter candidate, $\chi$, is produced in the early universe either through the freeze-out or freeze-in mechanism, depending on whether or not it has achieved equilibrium with the  thermal bath species, respectively. The processes that change the number density of DM are $\zp$ decays and annihilations through t and u-channels, and SM fermion annihilations through s-channel, as well as their backreactions in the case of freeze-out. These are illustrated in \autoref{fig:feynman_diagrams}.
In order to determine the DM relic density today, we must solve the Boltzmann fluid equation for its number density, $n_\chi$. We solve it in terms of the DM yield $Y_{\chi} = n_\chi/s$, with $s$ the entropy density. Taking into account the leading interactions in our model, we have,
\begin{equation}
\frac{sHx}{g_s^*(x)}\frac{dY_\chi}{dx} = \left[1-\left(\frac{Y_\chi}{Y_\chi^{eq}(x)}\right)^2\right] (\gamma_{\bar f f \to \bar \chi \chi}(x) + \gamma_{\zp \to \bar \chi \chi}(x) + \gamma_{\zp \zp \to \bar \chi \chi}(x) )\,,
\label{eq:boltz}
\end{equation}
where $x \equiv m_\zp/T$, $T$ is the temperature of the thermal bath, $H$ is the Hubble rate, $g_s^*(x)=1+\frac{1}{3}\frac{d\log g_s(T)}{d\log T}$, with $g_s(T)$ the number of effective degrees of freedom associated with entropy, $Y_\chi^{eq}$ is the DM yield at equilibrium, and the $\gamma$'s are the reaction rate densities for each process, given below\footnote{Note that at the resonance (where $T\approx m_{Z'}$) the s-channel process is essentially a $Z'$ decay to DM where the $Z'$ is produced through fermion annihilations.  Therefore, when $m_\zp > 2m_\chi$ we do not include $\gamma_{f\bar{f} \to \chi \bar\chi}$ at the resonance since $\gamma_{Z' \to \chi \bar\chi}$ already accounts for it.}.

The contribution of $\zp$ decays and annihilations to the production of $\chi$ depends strongly on whether they are part of the SM thermal bath, as otherwise they would be underabundant compared to the SM fermions.

For the decay process, we have
\begin{equation}\begin{split}
\gamma_{\zp \to \bar \chi \chi} (x) &= \frac{K_1(x)}{K_2(x)} n_\zp^{eq}(x) \Gamma_{\zp \to \bar \chi \chi}  \\
&= \frac{m_\zp^3 T}{8\pi^3} K_1\left(\frac{m_\zp}{T}\right) \sqrt{1-\frac{4m_\chi^2}{m_\zp^2}}\left[V_\chi^2\left(1+\frac{2m_\chi^2}{m_\zp^2}\right) + A_\chi^2\left(1-\frac{4m_\chi^2}{m_\zp^2}\right)\right],
\end{split}
\label{eq:rate1to2}
\end{equation}
where $K_n$ is the modified Bessel function of the second kind of order $n$, $n_i^{eq}$ is the equilibrium number density of species $i$, and the decay rate of $\zp$ into DM $\chi$ is given by,
\begin{align}
\Gamma_{\zp \to \bar \chi \chi} = \frac{m_\zp}{12\pi} \sqrt{1-\frac{4 m_\chi^2}{m_\zp^2}} \left[V_\chi^2\left(1+\frac{2m_\chi^2}{m_\zp^2}\right) + A_\chi^2\left(1-\frac{4m_\chi^2}{m_\zp^2}\right)\right]\,.
\label{eq:decayrate}
\end{align}

For the self-annihilation processes of bath species $b$, we have
\begin{equation}\begin{split}
\gamma_{b b \to \bar \chi \chi} (x) &= (n_b^{eq}(x))^2 \langle \sigma v \rangle_{b b \to \bar \chi \chi} = (n_\chi^{eq}(x))^2 \langle \sigma v \rangle_{\bar \chi \chi \to bb} \\
&\approx \sum_b \frac{1}{32(2\pi)^6}T \int ds \sqrt{s} K_1\left(\frac{\sqrt{s}}{T}\right) \sqrt{1-\frac{4m_\chi^2}{s}} \sqrt{1-\frac{4m_b^2}{s}} \int d\Omega_{13} |\M|_{b b \to \bar \chi \chi}^2   \,
\end{split}
\label{eq:rate2to2}
\end{equation}
where $\langle \sigma v \rangle$ are the thermally averaged cross sections.
In the expression above, the approximation holds for initial states obeying Maxwell-Boltzmann statistics, and $|\M|^2$ is the non-averaged squared amplitude. For the $Z'Z' \to \bar \chi \chi$ process, at high temperatures ($T>>m_\zp$), we have $\gamma_{Z'Z'\to \bar \chi\chi} \propto T^6$. The s-channel rates exhibit a resonance regime, in which we can use the narrow width approximation, and off-shell regimes for light $\zp$ ($m_\zp \ll T,\sqrt{s}$) and heavy $\zp$ ($m_\zp \gg T,\sqrt{s}$). The explicit expressions for the rate densities of our $2\to 2$ processes, as well as approximations, can be found in Ref. \cite{RoseGray:2020ltu}.

Roughly speaking, if all the reaction rates are always smaller than the cosmic expansion rate ($\gamma/n_b \ll H$), $\chi$ would not be able to thermalize and freeze-in production is then possible. 
Otherwise, freeze-out would take place, as $n_\chi$ becomes comparable to its equilibrium value and the backreactions begin to be relevant in \autoref{eq:boltz}.

\begin{figure}[!t]
\centering
\includegraphics{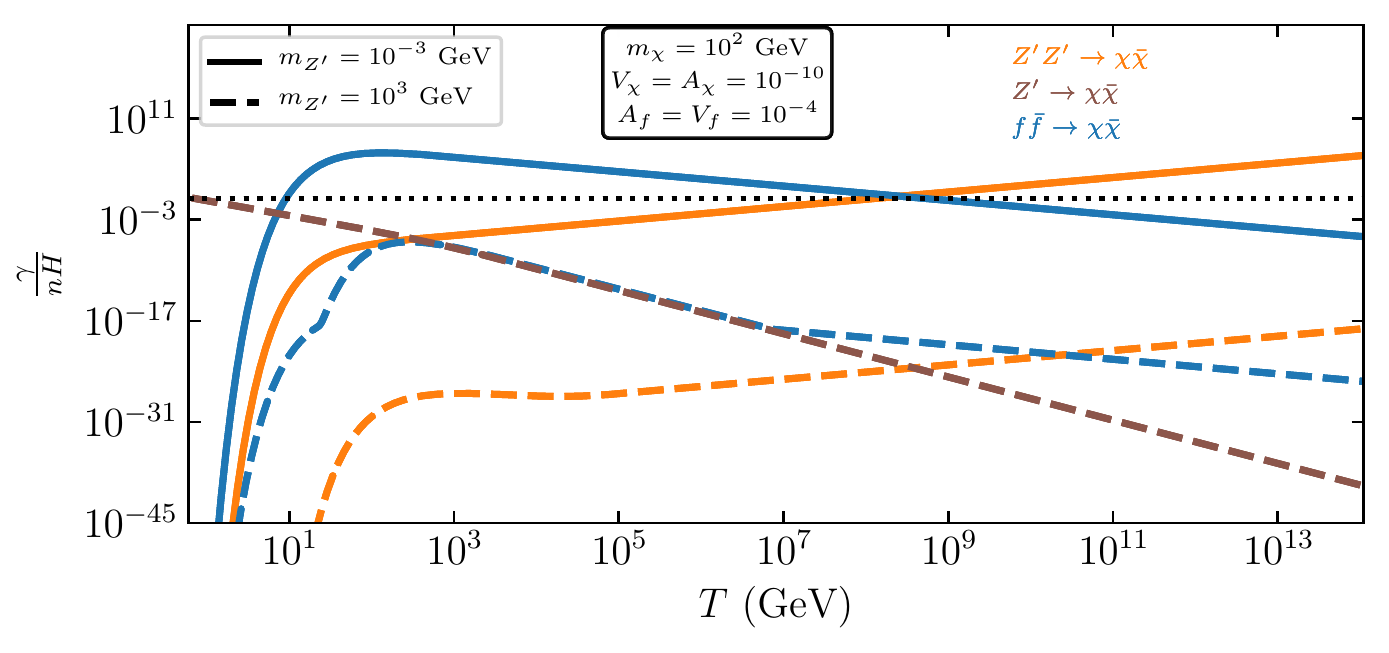}
\caption{Ratios of the production rates ($\gamma/n$) to the Hubble rate ($H$) for the three DM production processes: $\zp$ annihilations (in orange) and decays (in brown), and $f$ annihilations (in blue). The dotted black line shows where $\gamma/n=H$. For this set of parameters, we can see that the freeze-in regime is achieved for $m_\zp = 10^3$ GeV (dashed curves), whereas freeze-out occurs for $m_\zp = 10^{-3}$ GeV (solid curves). 
}
\label{fig:rates}
\end{figure}

In \autoref{fig:rates}, we show the ratio of the reaction rates 
to the Hubble rate as a function of $T$, for the $\zp$ decays (in brown) and annihilations (in orange), and SM fermion annihilations (in blue). The dotted horizontal line, where $\gamma/n_b = H$, indicates where we would have a rough boundary between freeze-in and freeze-out. Of course, the stronger the overall coupling, which is a function of the vector and axial couplings and the masses, the easier it is for $\chi$ to be produced via freeze-out. We illustrate the impact of $m_\zp$ on the production regime by considering it to be in the MeV scale (solid curves) and in the TeV scale (dashed curves), for a given set of couplings and DM mass as indicated in the figure. Note that a solid brown curve is not present since the decay is not kinematically allowed. As we can see, even for tiny couplings between $\chi$ and $\zp$ ($V_\chi = A_\chi = 10^{-10}$), the exchange of a light enough $\zp$ is able to sufficiently enhance the s-channel cross section to thermalize DM. On the other hand, the $\zp$ does not need to be much heavier than $\chi$ to suppress the rates and enable the freeze-in regime.

For each point in our parameter space, comprised of $m_\chi, m_\zp, V_\chi, A_\chi, V_f,$ and $A_f$ (assuming universal couplings to SM fermions), we check whether the sum of all the kinematically available processes are sub-Hubble for temperatures above their Boltzmann suppression. We therefore have in the next figures dashed curves indicating the \textit{regime boundary}, between freeze-in and freeze-out. Furthermore, we do the same analysis for the processes which would thermalize $\zp$ with the SM thermal bath. The region in our parameter space where $\zp$'s are not part of the SM bath will be indicated in blue and labeled "non-thermal $\zp$".

\begin{figure}[!t]
\centering
\includegraphics{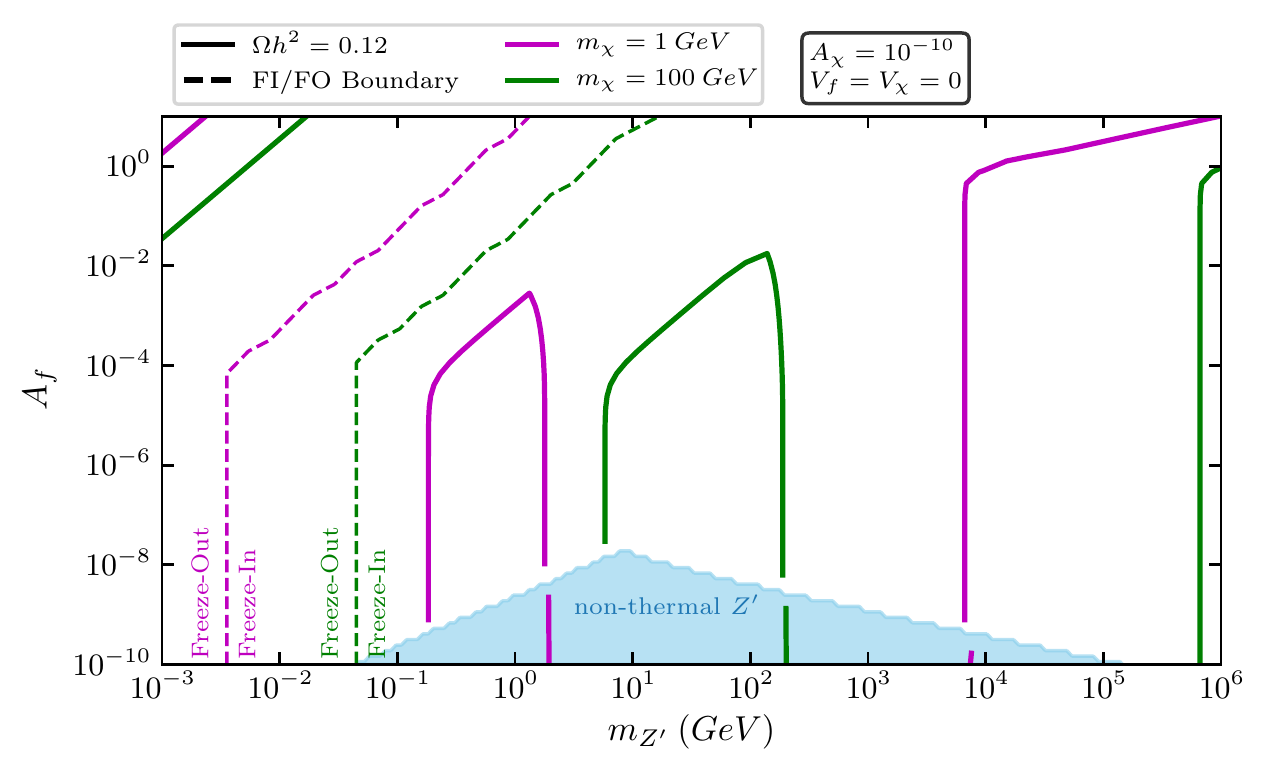}
\caption{Observed relic abundance contours for an axial $\zp$, where $\Omega_\chi^0 h^2 \simeq 0.12$ \cite{Aghanim:2018eyx} (solid curves), and regime boundaries (dashed curves), on $A_f$ vs $m_\zp$ plane.
Two DM masses are shown where $m_{\chi} = 1$ GeV is drawn in magenta, and $m_{\chi} = 100$ GeV in green. DM particles are regarded as WIMPs (FIMPs) in the region to the left (right) of the regime boundary. In the blue region, $\zp$ is not part of the SM thermal bath.}
\label{fig:contours}
\end{figure}

In \autoref{fig:contours}, we show in the plane ($m_\zp, A_f$) the contours of observed relic density of $\chi$ as inferred by Planck, $\Omega_\chi^0 h^2 \simeq 0.12$ \cite{Aghanim:2018eyx} (solid curves), produced either by freeze-out or freeze-in, to the left and to the right of the regime boundary, respectively. The regions between the freeze-out and freeze-in contours would lead to an overproduction of dark matter, which would overclose the universe, and are therefore excluded by Planck. As we have said, the purely vector $\zp$ case is extensively explored in the literature, so in this figure we show how different $m_\chi$ values impact the contours in the purely axial $\zp$ case. In magenta (green) we see the DM relic density contours as solid curves and the regime boundaries as dashed curves, for $m_\chi = 1$ GeV ($m_\chi = 100$ GeV). Also, we have chosen a small value for the dark matter coupling ($A_\chi = 10^{-10}$) in order to have a better understanding of the contours in the freeze-in region. 

As expected, we see in \autoref{fig:contours} that the region of our parameter space in which the $\zp$ is not part of the SM thermal bath (blue region) corresponds to small couplings, in this case, $A_f$. However, the thermalization between the $\zp$ and $f$'s depends strongly on $m_\zp$. The annihilations $\bar f f \to \zp \zp$ dominate the $\zp$ thermalization for lower masses, with rates increasing rapidly as the $\zp$ becomes lighter. In contrast, inverse decays $\bar f f \to \zp$ dominate for larger masses, with rates directly proportional to $m_\zp$. In the region of a non-thermal $\zp$, only the s-channels contribute to the DM relic abundance, since $n_\zp \ll n_\zp^{eq}\sim n_f^{eq}$
\footnote{In order to properly deal with the transition into the non-thermal $\zp$ region, we should solve the coupled Boltzmann equations for $n_\zp$ and $n_\chi$. However, as we will see in the next section, our goal is to study the phenomenology of our model, which happens to be in the region where $\zp$ can be safely regarded as thermal.}. In this case, the observed DM abundance can only be achieved for $m_\zp > 2m_\chi$.

Let us now focus on the regime boundary, indicated by dashed curves in \autoref{fig:contours} (as well as in \autoref{fig:limits}). As stated above, this boundary lies where the sum of the reaction rates of the processes depicted in \autoref{fig:feynman_diagrams} equals the Hubble rate. $\zp$ annihilations, which are independent of the SM couplings, dominate when $m_\zp < m_\chi$. We therefore recognize that the vertical part of the regime boundaries is due to t/u-channels. For larger values of SM couplings, the SM fermion annihilations start dominating the thermalization of $\chi$, corresponding to the diagonal parts of the regime boundaries. Decays and inverse decays can only dominate the $\chi$ thermalization for much heavier $\zp$'s, not considered in the parameter space of interest in this work. 

As we can see in \autoref{fig:contours}, the lighter the DM, the lighter the $\zp$ must be to provide the observed DM abundance. Accordingly, the regime boundary also shifts with the DM mass. The features of the freeze-out and freeze-in contours are discussed in the following subsections.

\subsection{Freeze-out regime}
\label{subsec:FO}

In the freeze-out regime, dark matter was initially part of the thermal bath, so the initial condition for \autoref{eq:boltz} is $Y_\chi = Y_\chi^{eq}$. The final relic density is found by using the usual freeze-out approximation \cite{Gondolo}.

In \autoref{fig:contours}, the contours providing the observed abundance of DM today are seen in the left upper corner (solid curves above the regime boundaries). In this case, WIMPs annihilate predominantly into SM fermions. As we will see later, though, WIMP annihilation into $\zp$'s dominate for smaller values of $m_\zp$. Since the case of WIMP dark matter with a $\zp$ portal is well-known, in \autoref{fig:contours} we choose to focus on the behavior of the FIMP relic density contours. Had we chosen much larger values of $A_\chi$ (see for instance Ref. \cite{Alves:2016cqf}), as we do in \autoref{sec:Results}, the parameter space providing viable WIMPs would be larger.

\subsection{Freeze-in regime}
\label{subsec:FI}

In the freeze-in regime, dark matter is assumed to be initially absent in the early universe, so the initial condition for \autoref{eq:boltz} is $Y_\chi = 0$. Since $Y_\chi$ is always much smaller than $Y_\chi^{eq}$ at least until the production finishes, we can safely neglect the term with $Y_\chi^2$ in the right-hand side of \autoref{eq:boltz}. We therefore simply integrate \autoref{eq:boltz} from the reheat temperature $T_R$, which we take to be $10^{14}$ GeV for the entire analysis, down to the current temperature. 
% \md{\textbf{I think it is better to say that in the next paragraph, where we introduce the distinction between IR and UV freeze-in. Dedicated bullets for each dominance regime (t/u-channel, s-channel light, resonance/decay, and s-channel heavy) might be a more clear way of explaining the different features. I'll try something like that ;-)}} \tg{I like this idea :)}

The freeze-in process can finish at the lowest scale available (infrared freeze-in), just like the usual case of freeze-out, or at the highest scale (ultraviolet freeze-in), in which case the relic abundance depends on the reheat temperature. Processes whose production rate densities have a high enough temperature-dependence can lead to ultraviolet freeze-in (as in the non-resonant $\zp$ portal of Ref.~\cite{Bhattacharyya:2018evo}). In the radiation era, this happens if the main process has $\gamma \propto T^n$ with  
% \cite{Biswas:2019iqm} 
$n>5$. Ultraviolet freeze-in is achieved via higher dimensional operators. In our model, we have two of such processes: the heavy $Z'$ regime of the s-channel, which happens as a four-fermion interaction and features $\gamma_{\bar f f \to \bar \chi \chi}\propto T^8/m_{Z'}^4$, and the t/u-channel annihilation of $Z'$ at high temperature in the presence of axial couplings (see discussion in \autoref{sec:UNIconstraints}), which features $\gamma_{\zp \zp \to \bar \chi \chi} \propto A_\chi^4 T^6 m_\chi^2/m_{Z'}^4$.
% In our case, $\gamma_{\zp \zp \to \bar \chi \chi} \propto T^6$ for $T\gg m_\zp$. 
Therefore, in regions of the parameter space where
% such 
these processes
% $\zp$ annihilations 
dominate, all FIMPs were produced around $T_R$. The $Z' \to \bar \chi \chi$ decay process freezes-in at $T\approx m_\zp$, when the number density of $Z'$ becomes Boltzmann-suppressed. The $\bar ff \to \bar \chi \chi$ process also freezes-in at $T \approx m_\zp$ whenever the resonance is allowed. In the case that $m_\zp < \max[m_{\chi},m_f]$, the process becomes Boltzmann-suppressed before the resonance occurs, therefore freeze-in happens at the smallest scale kinematically available, at $T\approx \max[m_{\chi},m_f]$. 

The contours of observed abundance produced through freeze-in shown in \autoref{fig:contours}
have interesting features due to the different production processes dominating freeze-in, as described below.

\paragraph{\underline{$\zp$ annihilation}} 

This process is only relevant for the achievement of the correct relic density in the presence of axial couplings, in which case it leads to UV freeze-in. In the high energy limit, when the $Z'$ momentum is much higher than its mass, the contribution of $Z'$ annihilations to the relic density is given by
\begin{equation}
    \Omega_\chi^0 h^2|_{t-ch} \sim 0.12 \left(\frac{100}{g_{eff}}\right)^{3/2} \left(\frac{m_\chi}{1 \text{GeV}}\right)^3 \left(\frac{0.19 \text{GeV}}{m_\zp}\right)^4 \left(\frac{A_\chi}{10^{-10}}\right)^4 \left(\frac{T_R}{10^{14} \text{GeV}}\right)\,,
\end{equation}
where for simplicity we set all degrees of freedom constant, $g_{eff}\equiv g_s = g_e$. For $m_\chi = 1$ GeV ($m_\chi = 100$ GeV), the t/u-channel sets the correct relic density for $m_\zp \sim 0.2$ GeV ($m_\zp \sim 5$ GeV).

\paragraph{\underline{$f$ annihilation in the light $Z'$ regime}}

For larger values of $A_f$, the s-channels begin to dominate, changing the slope of the contours. While $m_\zp < 2m_\chi$, the s-channels are in the light $Z'$ regime ($s \ll m_\zp^2$) and their contribution to the relic density is found to be 
\begin{equation}\label{eq:relicLIGHT}
    \Omega_\chi^0 h^2|_{light} \sim \frac{1.8\times 10^{25}}{g_{eff}^{3/2}}\sum_f \frac{m_\chi}{\text{max}(m_f,m_\chi)} \left[V_f^2(V_\chi^2+A_\chi^2)+A_f^2\left(V_\chi^2+A_\chi^2\left(1+\frac{12 m_\chi^2 m_f^2}{m_\zp^4}\right)\right)\right]
\end{equation}
As we can see from this expression, this process leads to a relic contour independent of $m_\zp$ in the absence of axial couplings (see top right panel of \autoref{fig:limits} below).

\paragraph{\underline{$f$ annihilation in the resonant $Z'$ regime / $Z'$ decay}}

When $m_\zp > 2 m_{\chi}$, the on-shell production of $\zp$, and the subsequent $Z'$ decay into dark matter, becomes possible. As a consequence, much smaller values of $A_f$ are required in order to not overproduce dark matter. In this case, we can use the narrow width approximation in \autoref{eq:rate2to2} to find the contribution of this process to the relic density. In the case of a purely axial $Z'$, as in \autoref{fig:contours}, we have
\begin{equation}
\Omega_\chi^0 h^2|_{\it{resonant}} \sim \frac{8.8\times 10^{25}}{g_{eff}^{3/2}} \frac{r_\chi A_\chi^2 \sqrt{1-4r_\chi^2} \sum_f A_f^2 \sqrt{1-4r_f^2}(1-4r_f^2-4r_\chi^2+16r_f^2 r_\chi^2)}{A_\chi^2(1-4r_\chi^2)\sqrt{1-4r_\chi^2}+\sum_f A_f^2(1-4r_f^2)\sqrt{1-4r_f^2}}\,,
\end{equation}
with $r_i \equiv m_i/m_\zp$.

As we can see, when $A_f^2 \gtrsim A_\chi^2$, as in \autoref{fig:contours}, the freeze-in contour in the $Z'$ resonance region is mostly independent of $A_f$. For the parameters chosen in \autoref{fig:contours}, the correct relic density in the $Z'$ resonance regime is found for $m_\zp \sim 2$ GeV ($m_\zp \sim 200$ GeV) and $m_\zp \sim 7.3 \times 10^3$ GeV ($m_\zp \sim 7.3 \times 10^5$ GeV) when $m_\chi = 1$ GeV ($m_\chi = 100$ GeV).

Note that the resonance is able to dominate the freeze-in contours even for $m_\zp \gg m_\chi$, in contrast to what happens in the freeze-out contours, with narrow resonances centered at $m_\zp \sim 2m_\chi$. This is because in the freeze-out case we integrate the annihilation cross-sections ($= \gamma_{\bar ff \to \bar \chi \chi}/(n_\chi^{eq})^2$, see \autoref{eq:rate2to2}) up to the freeze-out temperature ($T_f \approx m_\chi/30$). In the case of freeze-in, however, we integrate the production cross-sections ($= \gamma_{\bar ff \to \bar \chi \chi}/n_f^2$) up to the reheating temperature ($T_R \gg m_\chi$). 

\paragraph{\underline{$f$ annihilation in the heavy $Z'$ regime}}

When $m_\zp \gg m_\chi, m_f$, the s-channel happens as a four-fermion interaction, in the heavy $Z'$ regime. In the limit $m_\zp^2 \gg s \gg m_f^2,m_\chi^2$, with $m_\zp < T_R$, the contribution of the heavy regime is found by integrating the production rate over temperature up to $m_\zp$. We find
\begin{equation}\begin{split}
    \Omega_\chi^0 h^2|_{heavy} \sim \frac{4.0\times 10^{25}}{g_{eff}^{3/2}} \sum_f  (V_f^2 + A_f^2) (V_\chi^2+A_\chi^2) \frac{m_\chi}{m_\zp}\,.
\end{split}\end{equation}

If $m_\zp > T_R$, the contribution of the heavy regime to the relic density would instead depend on $m_\chi T_R^3/m_\zp^4$. In \autoref{fig:contours}, the heavy $Z'$ regime dominates the freeze-in contour for $m_\zp > 10^4$ GeV ($m_\zp > 10^6$ GeV) for $m_\chi = 1$ GeV ($m_\chi = 100$ GeV).

\section{Constraints on parameter space}
\label{sect:constraints}

In this section, we discuss the most stringent constraints on our parameter space. As we will see in the next section, they provide complementary bounds on the $\zp$ mass and couplings.

\textit{Indirect detection} searches for dark matter could also pose limits on our parameter space. However, since they are only sensitive to annihilation cross-sections near the thermal region, they would only constrain our WIMP scenario. Such constraints would not be competitive with the experimental constraints on $\zp$'s discussed below and are not considered here.

New species thermalized with the SM plasma at temperatures in the MeV scale can change the predictions of \textit{Big Bang Nucleosynthesis} (BBN) \cite{Sabti:2019mhn,Blanco:2019hah}. In this work, we restrict ourselves to the case of dark matter candidates at and above the GeV scale. Therefore, one should only be concerned about the lower bound on $\zp$ masses. Interactions like $e^+e^- \tob \nu \nu$, through the exchange of a light enough $\zp$, can delay the neutrino decoupling. While a detailed analysis of such effects is beyond the scope of this work and would not change our main conclusions, we adopt the conservative bound of $m_\zp > 10$ MeV. 

\subsection{Direct detection}
\label{sec:DDconstraints}

Direct detection experiments aim to identify the nuclear or electronic responses produced by the collisions between DM and the detector's target nuclei, being
able to place stringent constraints and/or rule out DM models. 

In our model, the scattering off nuclei takes place through t-channel exchanges of a $\zp$. When $m_\zp \lesssim \sqrt{2m_N E_R}$, with $m_N$ the nucleus mass and $E_R$ the recoil energy, the usual approximation of a short-range interaction via heavy mediators does not hold. In order to consider direct detection bounds with a sub-GeV $\zp$, we use the recasted limits provided by the micrOMEGAs package \cite{Belanger:2020gnr}. We will show in \autoref{fig:limits} the constraints from XENON1T \cite{XENON1T} on the spin-independent DM-nucleon scattering cross section, which provides the most sensitive current direct detection limits on our parameter space. 

\subsection{Experimental Constraints on $Z'$ Parameters}
\label{sec:contraints}

There are constraints on $Z'$ properties from a broad range of existing experiments ranging from low energy atomic parity violation measurements \cite{Diener:2011jt} to high energy searches at the LHC \cite{Aad:2019fac}. In what follows, we briefly discuss the most stringent ones, referring the interested reader to the existing literature for details. Also,  our list of measurements is not exhaustive as we do not include constraints that are less restrictive than the ones we describe below.  
In addition, these bounds only apply to $Z'$'s that couple to leptons and therefore do not apply to leptophobic $Z'$'s.
Our results are summarized in~\autoref{fig:limits}. 

\paragraph{LHC}

$Z'$ bosons can be produced via Drell-Yan production, $pp\to l^+ l^- X$, where $X$ represents the beam fragment jets, in hadron colliders \cite{Barger:1986nn,Rosner:1986cv,Barger:1986hd,delAguila:1986klm,Capstick:1987uc,Dittmar:2003ir}, so that constraints can be put on $Z'$ parameters by comparing the predicted $Z'$ production cross sections for specific final states, $\sigma (pp\to Z') \times BR(Z' \to l^+ l^-)$, to experimental limits on these cross sections \cite{Capstick:1987uc,Dittmar:2003ir}. Both CMS \cite{Sirunyan:2019vgt} and ATLAS  \cite{Aad:2019fac}, and prior to this CDF \cite{Aaltonen:2008vx,Aaltonen:2008ah} and D0 \cite{Abazov:2010ti}, have obtained such limits for specific $Z'$ models. In this work, we use the 95\% confidence level experimental limits on the cross section to dilepton final states given by the ATLAS collaboration \cite{Aad:2019fac} based on LHC Run 2 at $\sqrt{s}=13$~TeV with total integrated luminosity of $L=139$fb$^{-1}$. To calculate the theoretical predictions for the cross sections we use the expressions given in Ref.~\cite{Godfrey:1987qz}, the LHAPDF set C10 parton distribution functions \cite{Buckley:2014ana,Guzzi:2011sv}, and include the 1-loop K-factors to account for NLO QCD corrections \cite{KubarAndre:1978uy,Altarelli:1978id}. NLO QCD and electroweak radiative corrections were included in the width calculations \cite{Kataev:1992dg}. To obtain a limit on the $Z'$ couplings for a given $Z'$ mass, $m_\zp$, we take the ATLAS limit on $\sigma (pp\to Z') \times BR(Z' \to l^+ l^-)$ and vary the couplings for the given $m_\zp$ until we obtain agreement between the predicted value and the ATLAS limit. We checked the reliability of our calculations by comparing our results with the limits on $m_\zp$ for some of the models in Ref.~\cite{Aad:2019fac}. The resulting excluded parameter space is shown in~\autoref{fig:limits}.

\paragraph{$\boldsymbol{e^+e^-}$ with LEP II data}

One can put constraints on $Z'$'s by looking for deviations from SM predictions due to the interference
with the $Z'$ in $e^+e^- \to f\bar{f}$.  The LEP experiments, ALEPH, DELPHI, L3 and OPAL, have 
summarized their measurements for $130\; \hbox{GeV} \leq \sqrt{s} \leq 207\; \hbox{GeV}$ 
in Ref.~\cite{Schael:2013ita}.  They give results
for $\sigma(e^+e^- \to \mu^+\mu^-)$, $\sigma(e^+e^- \to \tau^+\tau^-)$, $\sigma(e^+e^- \to hadrons)$,
$A_{FB}(\mu+\mu-)$, and $A_{FB}(\tau^+\tau^-)$, where $A_{FB}$ are forward-backward asymmetries. We 
use the  expressions given in \cite{Capstick:1987uc} to calculate the predicted values for these observables.
To reduce the theoretical uncertainties we use the ratio of the observable with the $Z'$ divided by the SM prediction
and compare the deviation from 1 to the experimental error.  We construct a $\chi^2$ summing over all the
measurements given in Ref.~\cite{Schael:2013ita}  for the observables and energy range given above
to find the 95\% C.L. limit on $m_\zp$. The resulting excluded parameter space is shown in~\autoref{fig:limits}.

\paragraph{BaBar from $\boldsymbol{e^+e^- \to \gamma Z'}$}

The BaBar experiment has placed limits on dark photon properties from the process $e^+e^- \to \gamma A'$, followed by $A'  \to e^+e^-, \; \mu^+\mu^-$, where $A'$ is the dark photon \cite{Lees:2014xha}.  
Expressions for this process are given in Ref.~\cite{Essig:2009nc}.
However, instead of attempting to  properly take into account experimental details, such as detector acceptances 
and efficiencies, we follow a more pragmatic approach by digitizing the BaBar results \cite{Lees:2014xha} and rescaling them to obtain the excluded region in~\autoref{fig:limits}.
To obtain these limits we recalculated the expression for $e^+e^- \to \gamma Z'$, which 
resembled the expression for $e^+e^- \to \gamma A'$ with the substitution 
$\epsilon^2 e^2 \to  (V_{f}^2 + A_{f}^2) $, where $\epsilon$ is the kinetic mixing term 
between the dark photon and the SM photon and $e$ is the electric charge. 
We note that BaBar reported
a more recent result in  $e^+e^- \to \gamma A'$, where the $A'$ 
decays to DM, so with
invisible decay products  \cite{Lees:2017lec}. Nevertheless, the limits from this latter process are more stringent
in only  a few small regions of the parameter space and, therefore, we only show limits 
for the case with $\mu^+ \mu^-$ in the final state \cite{Lees:2014xha}.

\paragraph{LHCb from $\boldsymbol{A' \to \mu^+ \mu^-}$}

The LHCb experiment has placed limits on dark photon properties from the search for dark photons
produced in $pp$ collisions at $\sqrt{s}=13$~TeV which subsequently decay to $\mu^+\mu^-$ pairs
\cite{Aaij:2019bvg,Aaij:2017rft}.  
Analogous to the previous paragraph, we rescale the LHCb limits using the substitution $\epsilon^2 e^2 \to  (V_{f}^2 + A_{f}^2) $.
LHCb considers two scenarios,  a prompt-like $A'$ search and a long-lived $A' \to \mu^+\mu^-$.
Obtaining limits from the prompt-like $A'$ search is straightforward as lifetime dependent systematic effects cancel \cite{Aaij:2019bvg,Aaij:2017rft} and a $Z'$ with universal couplings to all SM particles will not alter this situation.
For the long-lived $A'$ results we follow Ref.~\cite{Ilten:2018crw} and use the supplementary data
from Ref.~\cite{Aaij:2019bvg} to take into account lifetime dependent detector efficiencies.  The long-lived $A'$ case does not give rise to any additional constraints.

\paragraph{Atomic Parity Violation in $\boldsymbol{^{133}_{55}\hbox{Cs}}$}

Atomic parity violation (APV) measurements offer some of the most precise tests of the SM electroweak theory
and has been used to constrain various types of new physics including $Z'$'s 
\cite{London:1986dk,Langacker:1990jf,Mahanthappa:1991pw,RamseyMusolf:1999qk,Porsev:2009pr,Porsev:2010de,Diener:2011jt,Williams:2011qb} (see Ref.~\cite{Safronova:2017xyt}  for a recent review).  
To constrain $Z'$'s using precision measurements of APV in $^{133}_{55}\hbox{Cs}$ we use
the expressions given in Ref.~\cite{Diener:2011jt}
and the  measurements given in the PDG \cite{10.1093/ptep/ptaa104}.  It is straightforward
to find the maximum allowed value of the $Z'$ couplings for a given value of $m_\zp$ at 95\% C.L..
It is important to keep in mind that the weak charge $Q_W$  is proportional to
$A_e[(2Z+N) V_u + (2N+Z) V_d]$ where $Z$ is the number of protons, $N$ is the number of neutrons
and $e$, $u$ and $d$ refer to the electron and up and down quarks respectively
so that we cannot obtain any constraint for the case of purely vector or purely axial couplings to
SM fermions.  The resulting limits are shown in~\autoref{fig:limits}.

\paragraph{Neutrino-electron scattering: $\boldsymbol{\nu_\mu e^- \to \nu_\mu e^-}$ and $\boldsymbol{\bar{\nu}_\mu e^- \to \bar{\nu}_\mu e^-}$}

Neutrino-electron scattering is another process that can constrain $Z'$'s 
\cite{London:1986dk,Godfrey:1987uw,Capstick:1987uc,London:1987gt,Williams:2011qb}. 
We constrain the $Z'$ parameters by comparing the neutral current parameters given by the
PDG \cite{10.1093/ptep/ptaa104} to the expressions given in Ref.~\cite{Capstick:1987uc}
to obtain 95\% C.L. limits shown in ~\autoref{fig:limits}. 

\paragraph{$\boldsymbol{(g-2)_\mu}$ and $\boldsymbol{(g-2)_e}$ }

The anomalous magnetic moments of the electron and muon can constrain new physics via loop 
contributions \cite{Jegerlehner:2009ry,Freytsis:2009bh,Queiroz:2014zfa,Williams:2011qb}, in particular, due to $Z'$'s 
\cite{Alves:2015mua,Bodas:2021fsy}.  
We use the expressions given in Ref.~\cite{Alves:2015mua} to calculate the contribution to $(g-2)_{\mu (e)}$
from our universal $Z'$.  
For $(g-2)_\mu$, we compare this to the deviation between the average of the recent Fermi National Accelerator Laboratory muon $(g-2)$ measurement  \cite{Abi:2021gix} and the  Brookhaven National Laboratory experiment E821  measurement \cite{Bennett:2004pv,Bennett:2006fi},
and 
the SM prediction \cite{Aoyama:2020ynm} (see also the electroweak review in the PDG
for details \cite{10.1093/ptep/ptaa104}).  The experimental average is larger than the SM prediction so we constrain
the $Z'$ parameters to give agreement with the experimental average at 95\% C.L.. The fitted parameter values (shown in \autoref{fig:limits} as the green bands) are completely ruled out by the neutrino-electron scattering and APV bounds. Specific charge assignments for our $Z'$ portal \cite{Cadeddu:2021dqx,Amaral:2021rzw,Bodas:2021fsy,Allanach:2015gkd} or different new physics is therefore needed to explain the $(g-2)_\mu$ anomaly.
For $(g-2)_e$, as noted in Refs.
\cite{Pospelov:2008zw,Williams:2011qb}, there is a subtlety in that $(g-2)_e$ is the most precise measurement used to determine
the fine structure constant and, consequently, the best bound to determine $(g-2)_e$
comes from the next most precise experiment that measures $\alpha$,
and not from the errors from the  $(g-2)_e$ experiments themselves.  Following \cite{Pospelov:2008zw,Williams:2011qb},
we use  $\delta (g-2)_e < 1.59\times 10^{-10}$ to constrain the $Z'$ parameters.  
The resulting limits are shown in ~\autoref{fig:limits}. 

\paragraph{Electron beam dump}

We can constrain $Z'$ parameters using limits from 
electron beam dump experiments 
\cite{Bjorken:2009mm,Williams:2011qb}.  In these experiments, $Z'$'s are produced via 
a bremsstrahlung-like process ($e^- N \to e^- N \zp$) where the electron beam is stopped in the target with the beam products stopped by shielding. A detector looks for the decay products downstream. Thus, by comparing 
the expected event rate to the experimental limits, we can constrain the $Z'$ properties.  
To obtain limits, the couplings need to have  ``Goldilocks'' values, i.e.,  not too small and not too big. On the one hand, it needs to be large enough for the $Z'$'s to be produced in sufficient quantity but, on the other hand, if it is too large, the $Z'$ will decay too quickly for the decay products to escape the shielding. In addition, if the couplings are too small, the $Z'$ will decay beyond the detector.
We follow Ref.~\cite{Williams:2011qb}, which uses the thick target approximation of Ref.~\cite{Bjorken:2009mm}.  
One difference to note between a dark photon ($A'$) and our $Z'$ is that our $Z'$ can decay to neutrinos, implying that the decay width will be larger than that of an $A'$ with similar mass and couplings. 
The limits from the electron beam dump experiments SLAC E137 \cite{Bjorken:1988as} and SLAC E141 \cite{Riordan:1987aw} are shown in~\autoref{fig:limits}. We also considered limits from the Fermilab experiment E774 \cite{Bross:1989mp} but they are weaker than the BBN bound and are not shown on these plots.

\subsection{Unitarity bounds}
\label{sec:UNIconstraints}

Our simplified dark matter model can violate perturbative unitarity at high energies.

In the high-energy limit, the self-annihilation $\bar \chi \chi \to \bar \chi \chi$ through a (longitudinal) $\zp$ exchange happens independently of the vector coupling \cite{Kahlhoefer:2015bea}. The partial-wave unitarity condition in this case implies a lower bound on the $\zp$ mass:
\begin{equation}\label{eq:unitarity}
m_\zp \gtrsim \sqrt{\frac{2}{\pi}} A_\chi m_\chi .
\end{equation}

An analogous relation holds for $\bar f f \to \bar f f$, with top quark self-annihilation providing the strongest limit on the $(m_\zp,A_f)$ plane. However, such a limit is not more stringent than the experimental ones and in \autoref{fig:limits} we will only show the consequence of \autoref{eq:unitarity}. 

Unitarity bounds are of course of most concern in the freeze-out regime of our model, in which $A_\chi$ is sizable. As a consequence, allowing for axial couplings can render WIMP models in tension with DM overproduction. Moreover, partial wave unitarity can establish weaker but almost model-independent upper limits on WIMP masses \cite{Griest:1989wd,Smirnov:2019ngs}. 

Self-annihilations of $\chi$ into a longitudinal $\zp$ violate unitarity for $\sqrt{s}>\pi m_\zp^2/m_\chi/A_\chi^2$ \cite{Kahlhoefer:2015bea}, such that new particles must be introduced for the consistency of the model -- potentially impacting both the relic density calculation and the phenomenology.

It is interesting to notice that FIMP models are usually safe from the unitarity perspective and, most importantly, the requirement of smaller axial couplings would not overproduce FIMPs. Also, for the tiny couplings involved in the freeze-in regime, the $\zp$ and $\chi$ can safely be taken to be much lighter than the particles providing their masses while restoring unitarity, which is not usually the case for simplified WIMP models.

\section{Results}
\label{sec:Results}

\begin{figure}
    \centering
    \includegraphics[width=0.49\linewidth]{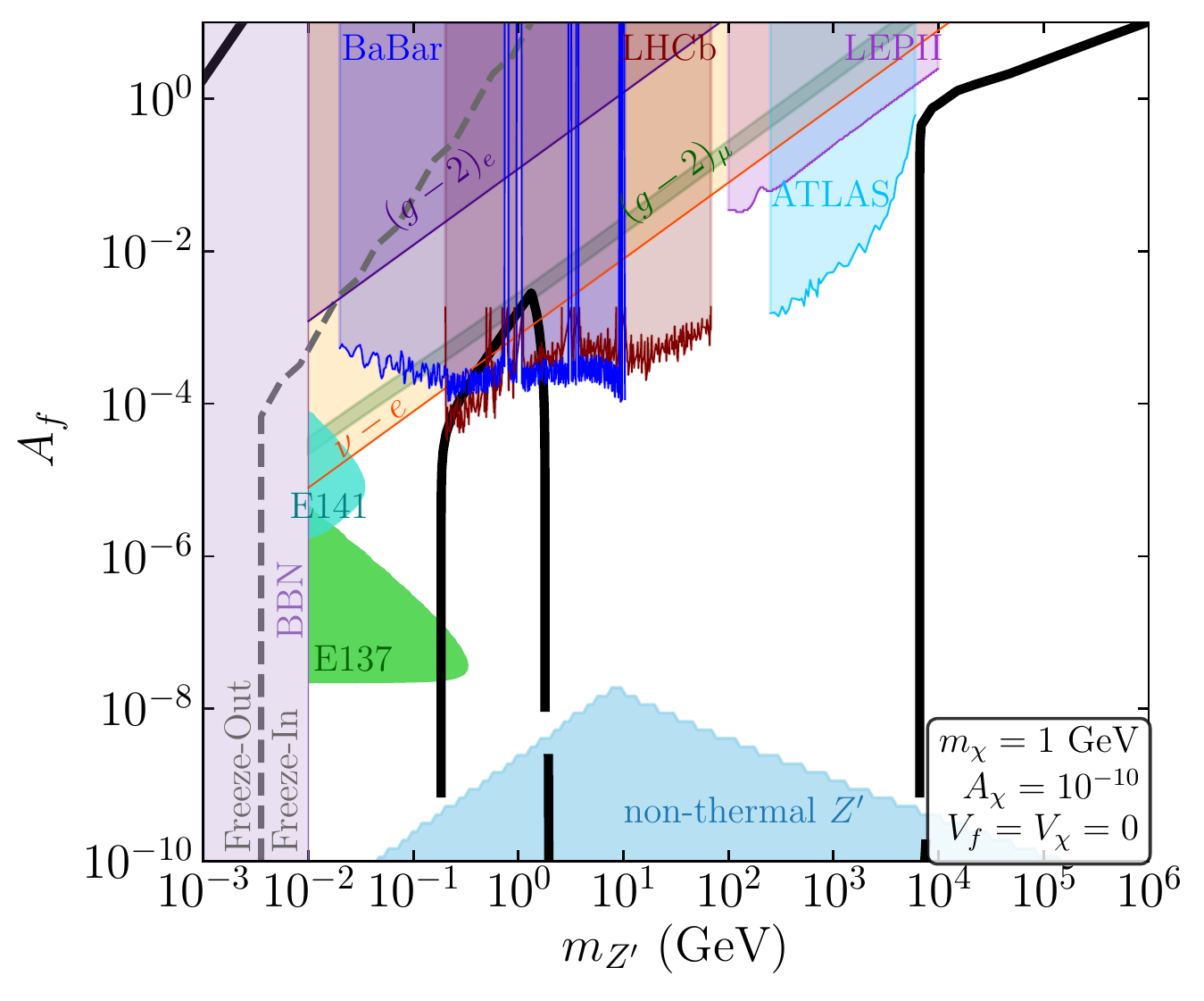}
    \includegraphics[width=0.49\linewidth]{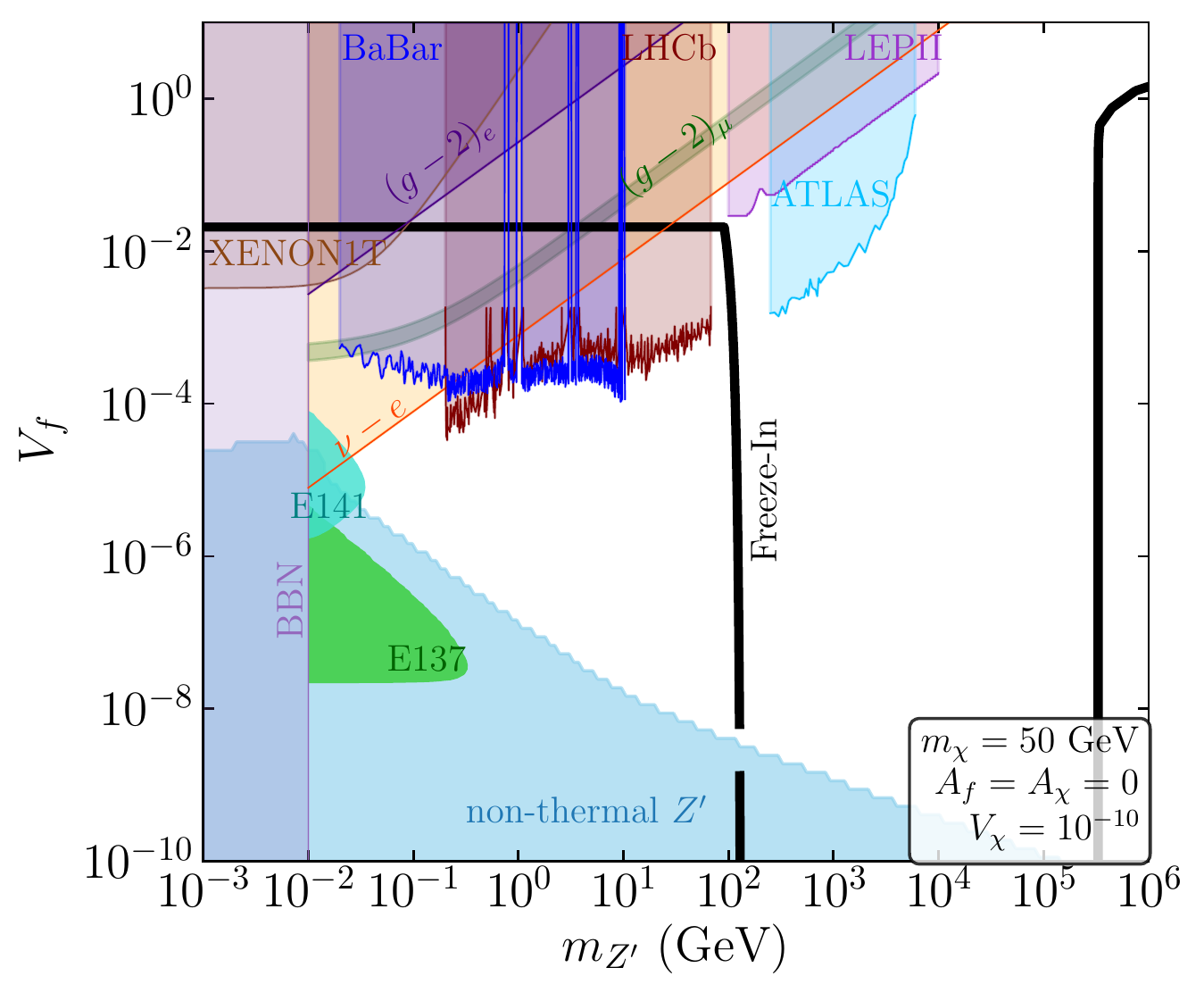}
    \includegraphics[width=0.49\linewidth]{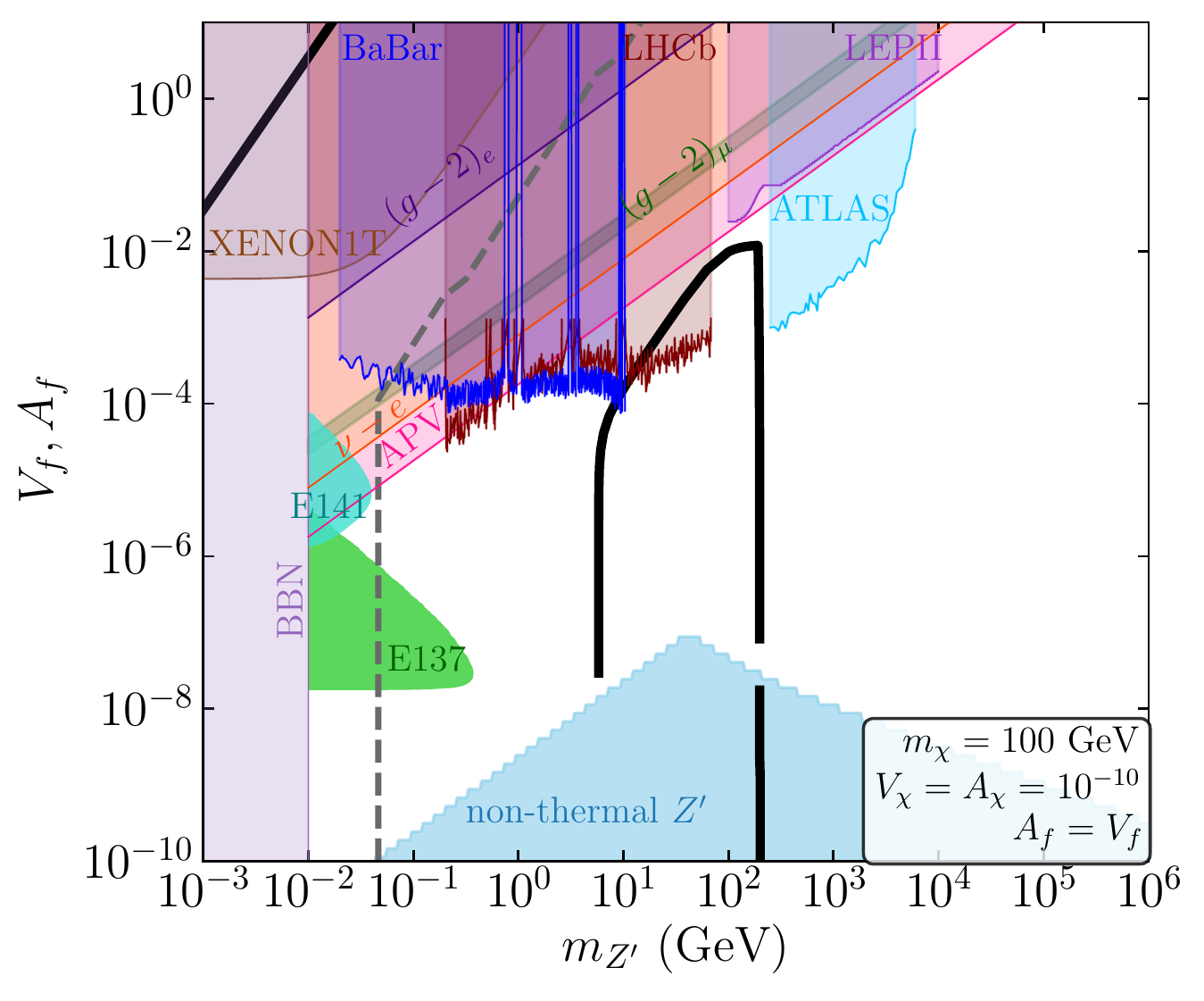}
    \includegraphics[width=0.49\linewidth]{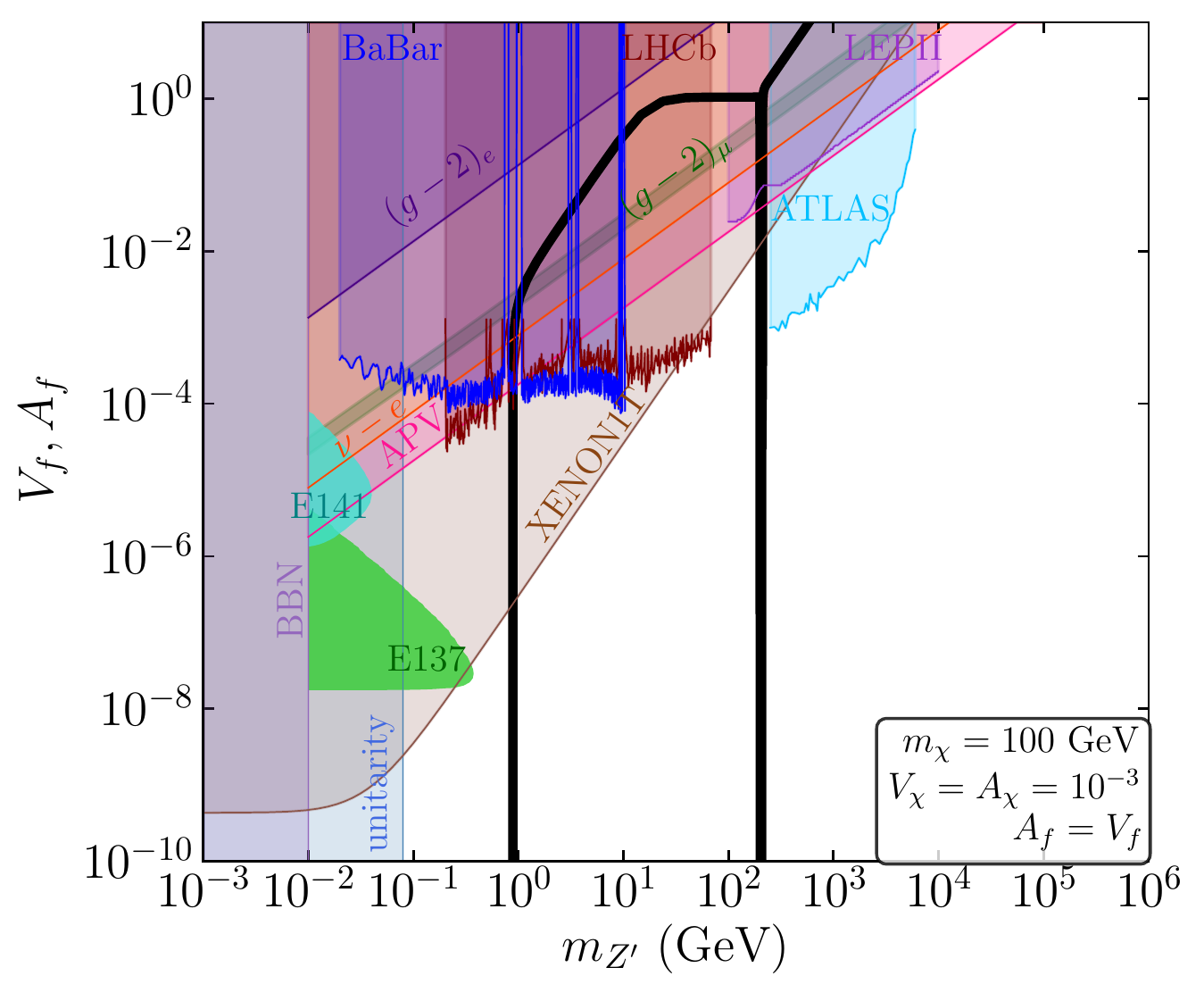}

    \caption{The relevant constraints (coloured shaded regions), each described in  \autoref{sect:constraints}, on
    the contours consistent with the observed DM abundance \cite{Aghanim:2018eyx} (solid black curves), for both freeze-out and freeze-in mechanisms. The dashed  grey curve corresponds to the regime boundary, with WIMPs (FIMPs) in the region to the left (right). In the region labeled as "non-thermal $\zp$" (in light blue), the $\zp$ was never thermalized with the SM particles. The panels on the top (bottom) right lack this regime boundary, hence in 
    the entire plane DM is produced through freeze-in (freeze-out).   The top left panel shows the scenario with purely axial couplings, for $m_\chi=1$ GeV, whereas the top right panel illustrates the case with purely vector couplings, for $m_\chi=50$ GeV. In the bottom panels, we present the scenario with both axial and vector couplings, for $m_\chi=100$ GeV, where $A_{\chi}=V_{\chi}=10^{-10}$ on the left and $A_{\chi}=V_{\chi}=10^{-3}$ on the right. 
    }
    \label{fig:limits}
\end{figure}

In this section, we present and discuss our results. In \autoref{fig:limits}, we show the constraints on our model, as described in \autoref{sect:constraints}, for different combinations of $A_{\chi/f}$ and $V_{\chi/f}$ and DM masses, in the plane of $\zp$ mass and SM-$Z'$ couplings. We show the contours providing the observed dark matter relic density produced via the freeze-out and freeze-in mechanisms (solid black curves) for a universal $\zp$ portal with purely axial couplings (top left panel), purely vector couplings (top right panel) and both axial and vector couplings (bottom panels). The dashed curves set the regime boundary, with DM being produced through freeze-out (freeze-in) in the region to the left (right) of the boundary. In the region labeled as ``non-thermal $\zp$'', the $\zp$ was never coupled to the SM fermions,
% part of the thermal bath, 
as discussed in \autoref{sec:relic density}. 

The top left panel of \autoref{fig:limits} illustrates the scenario where the $\zp$ couplings to both DM and SM fermions are purely axial, with $m_{\chi}=1$\,GeV and $A_{\chi}=10^{-10}$. This would be the case for a Majorana DM candidate. As shown in \autoref{fig:contours}, heavier DM would require a heavier  $Z'$ in order to agree with the relic abundance constraints. As one should expect, the freeze-out regime is completely excluded for such small $A_\chi$. 
Direct detection bounds on the spin-dependent DM scattering off nuclei, which would apply in this case, are much weaker than the other bounds
% too weak 
and therefore not relevant. 
However, independent experimental limits --
from neutrino-electron scattering, the beam dump experiment E137, and from the BaBar and LHCb collaborations -- are currently able to probe such an elusive frozen-in DM candidate. This remarkable result extends to DM masses in the range of $10$\,MeV-$10$\,TeV (with ATLAS and LEPII being able to probe the heavy DM limit and E141, the light one) and a wide range of $A_\chi$ (smaller values of $A_\chi$ require contours with larger values of $A_f$). Note that for larger values of $A_\chi$, $\chi$ is known to be a successful WIMP candidate which is able to evade the direct detection bounds (see for instance Ref.~\cite{Arcadi:2017kky}). 

In the top right panel of \autoref{fig:limits}, only vector couplings are assumed, with $V_{\chi}=10^{-10}$. As opposed to the previous case, in the absence of axial couplings the production rates are too weak to allow for thermal DM in this entire parameter space,
thus only the freeze-in mechanism is able to generate the observed DM abundance in this case.
We also observe that the $\zp$ is more easily decoupled from the SM fermions for the same reason\footnote{Note that processes like $G f \to Z' f$, with $G$ being SM gauge bosons, could still thermalize $Z'$, but our conclusions are unchanged.}. Moreover, the $Z'Z' \to \bar \chi \chi$ process is no longer as large near reheating, becoming negligible for the achievement of the relic abundance in this entire parameter space. In turn, the $\bar f f \to \bar \chi \chi$ process is relatively independent of $m_\zp$ if $m_\zp \lesssim \max[m_f,m_{\chi}]$ (or in other words, if the process becomes Boltzmann suppressed before the resonance) and $A_f = A_\chi = 0$ (cf. \autoref{eq:relicLIGHT}).
As it is already known \cite{Heeba:2019jho,Hambye:2018dpi}, spin-independent direct detection bounds on the DM-nuclei scattering provide strong constraints in this case, with freeze-in already being probed for DM masses from tens of GeV to a few TeV, provided that $m_\zp \ll m_\chi$. Similarly to the case of purely axial couplings, neutrino-electron scattering and bounds from the BaBar and LHCb collaborations put stringent constraints on the freeze-in contours. For larger values of $V_\chi$, smaller values of $V_f$ are needed and the freeze-in contours would also be constrained by direct detection and beam dump experiments (similarly to the case of a $Z'$ from $U(1)_{B-L}$ \cite{Heeba:2019jho}). Provided that $m_\zp<2m_\chi$, the conclusions above are independent of the DM mass. 

In the bottom left panel of \autoref{fig:limits}, we consider both axial and vector couplings of the $\zp$ to the SM fermions and DM, with a small value for the DM couplings, $V_\chi=A_\chi=10^{-10}$. Both freeze-out and freeze-in mechanisms can generate the observed DM abundance in the parameter space considered. For such small couplings to the  $\zp$, $\chi$ is ruled out as a WIMP by DM direct detection (limits from XENON1T) and BBN bounds, whilst LHCb and BaBar can constrain the FIMP scenario. Once again, setting the DM mass to smaller (larger) values would translate into a shift of the DM relic abundance curves to smaller (larger) $\zp$ masses. Since both axial and vector couplings are present, APV bounds are now possible and in fact are competitive with beam dump bounds for lighter DM candidates.

Finally, in the bottom right panel of \autoref{fig:limits}, we show the case where the $\zp$ couples both vectorially and axially to the SM fermions and DM, with relatively large DM couplings, $V_{\chi}=A_{\chi}=10^{-3}$. In such a scenario, DM was able to thermalize with the SM bath in the whole parameter space, therefore being produced via freeze-out. We can see that XENON1T limits rule out most of the parameter space providing the correct amount of WIMPs, except for the well known resonance region, where $m_{Z'}\approx 2m_{\chi}$ (see for instance \cite{Blanco:2019hah}) and the less explored light $\zp$ case, where $\bar \chi \chi \to Z' Z'$ annihilations dominate freeze-out production. As discussed in \autoref{sec:UNIconstraints}, for such large $\chi-\zp$ couplings, the unitarity lower bound on $m_\zp$ shown in \autoref{eq:unitarity}
becomes able to constrain part of our parameter space, as opposed to the previous cases. 
As a consequence, WIMPs cannot be too heavy in this case. Also note that, even though $\bar \chi \chi \to Z' Z'$ annihilations make WIMPs viable in a wide range of our parameter space, they violate unitarity at (not too high) energies. This renders such a simplified model less appealing, pointing towards the need of more realistic realizations (see for instance \cite{Kahlhoefer:2015bea}).

In summary, as evident by \autoref{fig:limits}, direct detection searches, experimental constraints on $\zp$, and unitarity and cosmological bounds can currently probe and/or exclude a significant part of our parameter space in a complementary way. We have shown that, even if $\zp$'s have very tiny couplings to dark matter and considerably small couplings to standard fermions, they are able to provide a successful freeze-in and are not invisible to current experimental searches, even in the case where they are purely axial.
By considering a purely vector $\zp$ (top right panel of \autoref{fig:limits}), larger values of $V_\chi$ are required in order for DM to be produced through freeze-out, as compared to the pure axial case.
Since in this case the correct DM abundance is produced when $A_f$ and $V_f$ are large enough, DM direct detection experiments can currently probe the freeze-in regime of the model. Future direct detection experiments such as DARWIN \cite{Aalbers_2016} and XENONnT \cite{Aprile_2020} are also not sensitive enough to probe freeze-in if axial couplings to DM exist.
We estimate that XENONnT limits probe couplings an order of magnitude smaller than the XENON1T limits for $100\:GeV$ dark matter, which still does not reach our freeze-in contours.
Nevertheless, even though direct detection fails to probe freeze-in in this case, the experimental constraints on $Z'$ parameters
that we have considered are able to significantly constrain freeze-in DM, while leaving viable a large part of our parameter space, particularly for smaller couplings and larger $m_{Z'}$. Moreover, one should note that, in our scenario, the regions below the freeze-out \textit{and} above the freeze-in contours are excluded by the Planck constraint, as they would overclose the universe.

\section{Conclusions}
\label{sec:conclusion}

In this work, we studied the freeze-out and freeze-in production of a Dirac fermion dark matter candidate, $\chi$, which interacts with the Standard Model fermions, $f$, via a universal $\zp$ portal. We have shown how the free parameters of our model, the vector and axial-vector couplings of $\zp$ ($V_f, A_f, V_\chi$, and $A_\chi$) and the masses of $\chi$ and $\zp$ ($m_\chi$ and $m_\zp$), impact the thermalization of $\chi$ and $\zp$ in the early universe. Our main results are presented in \autoref{fig:limits}.

We have discussed the role of each process (depicted in \autoref{fig:feynman_diagrams}) for the achievement of the observed relic abundance of $\chi$ (along the solid curves in \autoref{fig:contours} and \autoref{fig:limits}). We found that the t-channel can only dominate the correct abundance in the presence of axial couplings, setting a lower viable value of $m_\zp$ according to Planck, while s-channels dominate above this lower limit. In the absence of axial couplings (as shown in the top right panel of \autoref{fig:limits}), the relic contours are dominated by s-channels and are almost independent of $m_\zp$ when $m_\zp < 2 m_\chi$.

We explored the phenomenology of this model, considering a wide range of $\zp$ masses (from MeV up to PeV) and couplings. We considered DM direct detection bounds (XENON1T), experimental constraints on $\zp$ parameters (from colliders, neutrino-electron scattering, atomic parity violation, electron and muon anomalous magnetic moments, and beam dump experiments), as well as cosmological (BBN) and unitarity bounds,
as summarized in \autoref{fig:limits}. Our main result is that most of these constraints can already test freeze-in in a complementary way, with viable regions such as where $m_\zp \gg m_\chi$, mostly unconstrained.

As expected, if the $Z'$ has purely vector couplings (top right panel of \autoref{fig:limits}), existing XENON1T limits can exclude the freeze-in DM production for $m_\zp \ll m_\chi$, i.e., for a sub-GeV $\zp$. 
Additionally, constraints from BaBar, LHCb, and neutrino-electron scattering  measurements provide much stronger bounds in the case where $V_\chi \ll V_f$, making it possible to test heavier $Z'$'s and lighter $\chi$'s. Larger values of $V_\chi$ make electron beam dump experiments sensitive to freeze-in along with direct detection bounds. Since only s-channels are responsible for freeze-in production in this case, these conclusions hold for $m_\chi$ from tens of MeV up to tens of TeV provided that $m_\zp < 2 m_\chi$.

Weakening direct detection bounds by considering $V_\chi = 0$ (as for a Majorana dark matter candidate) is known to be one of the viable options for WIMPs in simplified models. In this work we focused instead on the FIMP regime of a purely axial $\zp$ (top left panel of \autoref{fig:limits}). While direct detection bounds are indeed too weak to be relevant in this case, the other experimental bounds on $A_f$ are still very stringent and able to rule out part of the viable FIMP parameter space, especially near the resonance ($m_\zp \sim 2 m_\chi$). We have therefore found that for a wide range of $A_\chi$ and $m_\chi$ in the range $100$~MeV-$100$~GeV, FIMPs which interact via a purely axial $\zp$ are also currently constrained by data.

In the presence of axial couplings, the production rates are stronger compared to the vector only case, making it easier for $\chi$ and $\zp$ to have thermalized with $f$ in the early universe. The scenario of both vector and axial couplings has more stringent constraints on both FIMPs and WIMPs, now coming from atomic parity violation. If the DM couplings are very small (bottom left panel of \autoref{fig:limits}), FIMPs can be tested mainly near the resonance region (for $m_\zp < 2m_{\chi}$), similarly to the case of pure axial $\zp$'s. Direct detection bounds are not strong enough to probe FIMPs though, since the freeze-in contours are no longer independent of $m_\zp$ for $m_\zp \ll m_\chi$ if axial couplings exist. Larger DM couplings (bottom right panel of \autoref{fig:limits}) make WIMPs viable DM candidates in very narrow regions near the $\zp$ resonance and at lower $m_\zp$ values set by the t-channel contribution. The unitarity bounds prevent WIMPs from being too heavy. Also, restoring unitarity violation due to t-channels would usually require the introduction of new states which cannot be too heavy, as opposed to the FIMP case.

In summary, the proposed model offers viable dark matter candidates whose experimental signatures can already be constrained by data from a variety of complementary search strategies, showing that part of the parameter space of both FIMPs and WIMPs mediated by a $\zp$ boson can be probed at present. Interestingly, we note that, although elusive, FIMP DM can currently be probed by a variety of experiments. This motivates further work on different realizations of our $\zp$ portal, as well as the development of even more sensitive searches for new feebly interacting particles.

\acknowledgments 

We would like to thank 
Genevi\`eve B\'elanger, Ali Mjallal, and Alexander Pukhov for their help with the recasting of direct detection bounds in micrOMEGAs, and Mike Williams, Nicol\'as Bernal, and Chee Sheng Fong for helpful communication. C.C. is supported by the Generalitat Valenciana Excellence grant PROMETEO-2019-083 and by the Spanish MINECO grant FPA2017-84543-P and the European Union’s Horizon 2020 research and innovation programme under the Marie Skłodowska-Curie grant agreement No 860881-HIDDeN. C.C. and M.D. were supported by the Arthur B. McDonald Canadian Astroparticle Physics Research Institute. This work was supported by the Natural Sciences and Engineering Research Council of Canada.

\appendix
\bibliographystyle{JHEP}
\bibliography{biblio}

\end{document}